\documentclass[%
pre,preprint,
tightenlines,
showpacs,showkeys,
twocolumn,
a4paper,11pt
]{revtex4}
\usepackage{dcolumn}
\usepackage{amssymb,amsmath,array}
\usepackage{graphicx}

\makeatletter

\newcommand{\bs}[1]{\boldsymbol{#1}}
\newcommand{\vc}[1]{\mathbf{#1}}
\newcommand{\uvc}[1]{\mathbf{\hat #1}}

\newcommand{\dd}{\mathrm{d}}
\newcommand{\pdr}[1]{\frac{\partial #1}{\partial t}}
\newcommand{\drf}[1]{\frac{\dd #1}{\dd t}}
\newcommand{\trans}{\emph{trans} }
\newcommand{\cis}{\emph{cis} }

\makeatother

\begin{document}
\DeclareGraphicsExtensions{.jpg,.pdf}
\title{Kinetics of photoinduced anisotropy
  in azopolymers:\\ models and mechanisms}

\author{A.D.~Kiselev}
\email[Email address: ]{kisel@elit.chernigov.ua}
\affiliation{%
 Chernigov State Technological University,
 Shevchenko Street 95,
 14027 Chernigov, Ukraine
} 

\date{\today}

\begin{abstract}
  We consider the effect of photoinduced optical anisotropy (POA) in
  azopolymers.  Using a unified approach to the kinetics of
  photo-reorientation we discuss the assumptions underlying the known
  theoretical models of POA and formulate a tractable phenomenological
  model in terms of angular redistribution probabilities and order
  parameter correlation functions.  The model takes into account
  biaxiality effects and long term stability of POA in azopolymers. It
  predicts that under certain conditions two different mechanisms,
  photoorientation and photoselection, will dominate POA depending on
  the wavelength of pumping light.  By using available experimental
  data, we employ the model to compute dependencies of principal
  absorption coefficients on the illumination time.  Our calculations
  clearly indicate the different regimes of POA and the numerical
  results are found to be in good agreement with the experimental
  data.

\end{abstract}

 
\maketitle

\section{Introduction}
\label{sec:intro}

The ability of some photosensitive materials to become dichroic and
birefringent under the action of light is known as the effect of
photoinduced optical anisotropy (POA).  This effect provides a
means of having the light-controlled anisotropy and it
makes the materials that
exhibit POA very promising for use in many
photonic applications~\cite{Pras:1995,Eich:1987}.
Owing to extremely high efficiency of POA in
side-chain polymers that contain covalently linked photo\-chro\-mic
moieties such as azobenzene derivatives and are known as azopolymers,
light induced ordering processes in these polymers have been
the subject of intense experimental and theoretical studies
in the last decade
~\cite{Eich:1987,Hvil:1992,Wies:1992,Dum2:1993,Stu:1994,Ped:1997,Ped:1998,Nat:1998,Puch:1998,Puch:1999}.

A typical experimental procedure to induce optical
anisotropy in azopolymers consists in irradiating a sample with polarized UV
light.  In this case the accepted though not very well understood
mechanism of POA assumes that the key processes involved are induced
\textit{trans--cis}-photo\-iso\-meri\-zation and subsequent thermal
and/or photochemical \textit{cis--trans}-back-iso\-meri\-zation of the
azobenzene moieties.  

These \textit{trans--cis--trans} photoisomerization cycles are
accompanied by rotations of the azobenzene chromophores.  Since the
transition dipole moment of the azobenzene moiety is directed along
its long molecular axis, the fragments oriented perpendicular to the
incident actinic light polarization vector, $\vc{E}$, are almost
inactive. The long axes of the azobenzene fragments tend to
become oriented along directions normal to the polarization vector
$\vc{E}$.  Non-photoactive groups then undergo reorientation due to
cooperative motion or dipole
interaction~\cite{Nat:1998,Puch:1998,Puch:1999}.
 
The above scenario assumes angular redistribution of the long axes of
the \trans molecules during the \textit{trans--cis--trans}
isomerization cycles.  It is known as the photoorientation mechanism
that was initially suggested in Ref.~\cite{Nep:1963} for the case,
where the lifetime of \cis state is short. This state becomes
temporary populated during photoisomerization but reacts immediately
back to thermodynamically stable \trans isomeric form.

Another limiting case~--~the so-called angular selective hole burning
mechanism (photoselection)~--~occurs when the \cis states are
long-living.  In this case POA is caused by the selective depletion of
the \trans isomeric form when reaching the steady
state~\cite{Dum2:1993}.  The anisotropy induced in this way is not
long term stable and disappears as a result of the thermal back
reaction.

In both cases the effect is primary governed by the dependence of
photoisomerization rates on orientation of the photoactive groups.
From the other hand, we have also seen that physical characteristics
of POA can be different depending on a number of additional factors
such as lifetime of the \cis form.  These factors will determine
kinetics of POA that describes how amount of photoinduced anisotropy
characterized by absorption dichroism or birefringence evolves in time
upon illumination and after switching it off.

In particular, as opposed to the reversible POA, where anisotropy
disappears after switching off the irradiation
~\cite{Nep:1963,Tod:1983,Dum:1992,Dum:1993,Dum:1996}, POA can be long
term stable.  This is the case for POA in liquid crystalline (LC)
azopolymers.  Theoretically, it means that the photo-reorientation in
LC azopolymers is a non-equilibrium process in a rather complex
polymer system and it still remains a challenge to develop a tractable
microscopic theory treating the effect adequately.

In this paper we present theoretical considerations concerning the
kinetics of POA by starting from kinetic rate equations of general
form and assuming that reorientation of the azobenzene groups results
in the appearance of a self-consistent anisotropic field that support
the photoinduced anisotropy.  This field is thought of as being caused
by anisotropic interactions between the azobenzene fragments and
rearrangement of the main chains and other non-absorbing fragments.

There are two phenomenological models based on similar assumptions:
the mean field model proposed in Ref.~\cite{Ped:1997,Ped:1998} and the
model with additional order parameter attributed to the polymer
backbone~\cite{Puch:1998,Puch:1999}.  
Both these models, though they look different, 
incorporate the long term stability by introducing additional
degree of freedom (subsystem) which kinetics reflects cooperative
motion and account for non-equilibrium behavior.

In Sec.~\ref{sec:model} we 
discuss physical assumptions underlying general structure of
the phenomenological models
and show that it is determined by the
angular redistribution probabilities and the order parameter
correlation functions.  We use the results of general analysis to
formulate a simple model of the photoinduced ordering in LC
azopolymers.  In addition to the long term stability of POA, this
model accounts for biaxiality of the photoinduced orientational
structures~\cite{Wies:1992,Kis:cond:2001,Yar:a:2001,Yar:2001,Kis:epj:2001,Hvil:2001}
that was neglected in the models of
Refs.~\cite{Ped:1997,Ped:1998,Puch:1998,Puch:1999}.

In Sec.~\ref{sec:num-res} we compute order parameter components and
fractions of the azobenzene units for different irradiation doses.
Then we use the results of numerical analysis to fit the experimental
data on photoinduced dichroism of absorption.  These data demonstrate
that for long-living \cis forms dependencies of the absorption
dichroism on the illumination time can be qualitatively different
depending the wavelength of exciting beam.  We find that the different
regimes can be explained by using our model and the calculated
dependencies are in good agreement with the data obtained
experimentally.

Finally, in Sec.~\ref{sec:concl} we draw together the results and make
some concluding remarks.  

\section{Model}
\label{sec:model}

We shall assume that the dye molecules in 
the ground state  are of \trans form 
and the orientation of the molecular axis is
defined by the unit vector 
$\uvc{n}=$($\sin\theta\cos\phi$,$\sin\theta\sin\phi$,
$\cos\theta$), where 
$\theta$ and $\phi$ are Euler angles of the unit vector.
Angular distribution of the \trans molecules at time $t$
is characterized by the number distribution function $N_{tr}(\uvc{n},t)$.
Similarly, molecules in the excited state have the \cis conformation and
are characterized by the function $N_{cis}(\uvc{n},t)$.
Then the number of \trans and \cis molecules is given by
\begin{gather}
  N_{tr}(t)\equiv N n_{tr}(t)=\int N_{tr}(\uvc{n},t)\,\dd\uvc{n},
  \label{eq:1t}\\
  N_{cis}(t)\equiv N n_{cis}(t)=\int N_{cis}(\uvc{n},t)\,\dd\uvc{n},
  \label{eq:1c}\\
n_{tr}(t)+n_{cis}(t)=1,
  \label{eq:consv}
\end{gather}
where 
$\displaystyle
\int\,\dd\uvc{n}\equiv 
\int_{0}^{2\pi}\dd\phi\int_{0}^{\pi}\,\sin\theta\,\dd\theta\,$ 
and 
$N$ is the total number of molecules.
The normalized angular
distribution functions, $f_\alpha(\uvc{n},t)$, of
\trans ($\alpha=tr$) and \cis ($\alpha=cis$) molecules 
can be conveniently defined by the relation
\begin{equation}
  \label{eq:dis}
  N_\alpha(\uvc{n},t)=N n_\alpha(t) f_\alpha(\uvc{n},t)\, . 
\end{equation}

We also need to introduce additional angular distribution function
$f_p(\uvc{n},t)$ characterizing the anisotropic field due to
interaction between a side chain fragment and nearby molecules.  In
particular, this field is affected by collective degrees of freedom of
non-absorbing units such as main chains and determines angular
distribution of the molecules in the stationary regime. It bears close
resemblance to the equilibrium distribution of the mean field theories
of POA. In Refs.~\cite{Ped:1997,Ped:1998,Hvil:2001} this distribution
has been assumed to be proportional to $\exp(-V(\uvc{n})/k_B T)$,
where $V(\uvc{n})$ is the mean-field potential that depends on the
order parameter tensor.

In other words, we have the additional subsystem characterized by
$f_p(\uvc{n},t)$ attributed to the presence of long-living angular
correlations coming from anisotropic interactions between side chain
groups and collective modes of polymeric environment.  For brevity, we
shall refer to the subsystem as a polymer system (matrix).  We shall
write the kinetic rate equations for $N_{\alpha}(\uvc{n},t)$ in the
form of master equations~\cite{Gard,Kamp}:

\begin{align}
&    
\pdr{N_\alpha}=\left[\drf{N_\alpha}\right]_{\mathrm{Diff}}+
\sum_{\beta\ne\alpha}\int
\Bigl[\,W(\alpha,\uvc{n}\,|\,\beta,\uvc{n}')\notag\\
&
\times N_\beta(\uvc{n}',t)-
W(\beta,\uvc{n}'\,|\,\alpha,\uvc{n})\,
N_\alpha(\uvc{n},t)\,
\Bigr]\,\dd\uvc{n}'\notag\\
&
+\gamma_{\alpha}
\Bigl[\,N_{\alpha}(t)
\int\Gamma_{\alpha-p}(\uvc{n},\uvc{n}') f_p(\uvc{n}',t)\dd\uvc{n}'
\notag\\
&
- N_{\alpha}(\uvc{n},t)\,\Bigr]\,,
\label{eq:master}
\end{align}
where $\alpha\,, \beta\in\{tr,\,cis\}$.

The first term on the right hand side of Eq.~\eqref{eq:master} is due to
rotational diffusion of molecules in \trans ($\alpha=tr$) and \cis 
($\alpha=cis$)
conformations. In frictionless models this term is absent and
will be dropped in our subsequent notations.

Now we need to specify the rate of 
$trans \to cis$ photoisomerization
stimulated by the incident UV light. 
For the electromagnetic 
wave linearly
polarized along the $x$--axis the transition rate
can be written as follows~\cite{Dum:1992,Dum:1996}:
\begin{align}
&
W(cis,\uvc{n}\,|\,tr,\uvc{n}')=
\Gamma_{t-c}(\uvc{n},\uvc{n}')\,P_{tr}(\uvc{n}'),
  \label{eq:wc-t}\\
&
P_{tr}(\uvc{n})=(\hbar\omega_t)^{-1}\Phi_{tr\to cis}\sum_{i,j}
\sigma_{ij}^{(tr)}(\uvc{n})E_i E_j^{*}\notag\\
&
=q_t I (1+u\, n_x^2)\,,
\label{eq:ptr}
\end{align}
where 
$\bs{\sigma}^{(tr)}(\uvc{n})$ is the tensor of 
absorption cross section
for the \trans molecule oriented along $\uvc{n}$:
${\sigma}^{(tr)}_{ij}=\sigma_{\perp}^{(tr)}\delta_{ij}+
(\sigma_{||}^{(tr)}-\sigma_{\perp}^{(tr)})\,
n_i\,n_j$;
$
u\equiv
(\sigma_{||}^{(tr)}-\sigma_{\perp}^{(tr)})/\sigma_{\perp}^{(tr)}
$
is the absorption anisotropy parameter;
$\hbar\omega_t$ is the photon energy;
$\Phi_{tr\to cis}$ is the quantum yield of the process and 
$\Gamma_{t-c}(\uvc{n},\uvc{n}')$ describes the angular redistribution of
the molecules excited in the \cis state;
$I$ is the pumping intensity and
$q_t\equiv (\hbar\omega_t)^{-1}\Phi_{tr\to cis}\sigma_{\perp}^{(tr)}$.
 
Similar line of reasoning applies
to the $cis\to trans$ transition, so we have
\begin{align}
  \label{eq:wt-c}
& W(tr,\uvc{n}\,|\,cis,\uvc{n}')=
\gamma_c\Gamma_{c-t}^{(sp)}(\uvc{n},\uvc{n}')\notag\\
&
+q_c I\,\Gamma_{c-t}^{(ind)\,}(\uvc{n},\uvc{n}')\,,
\end{align}
where 
$q_c\equiv (\hbar\omega_t)^{-1}\Phi_{cis\to trans}\sigma^{(cis)}$ and
$\gamma_c\equiv 1/\tau_c$, $\tau_c$ is the lifetime of \cis
molecule and the anisotropic part of the absorption cross section is
disregarded,
$\sigma_{||}^{(cis)}=\sigma_{\perp}^{(cis)}\equiv\sigma^{(cis)}$.
Eq.~\eqref{eq:wt-c} implies that the process of angular redistribution
for induced and spontaneous transitions can differ. 
Note that the normalization condition for all the angular redistribution
probability intensities is
\begin{equation}
  \label{eq:norm}
  \int\Gamma_{\beta-\alpha}(\uvc{n},\uvc{n}')\,\dd\uvc{n}=1\, .
\end{equation}

From Eqs.~\eqref{eq:master},~\eqref{eq:wc-t} and~\eqref{eq:wt-c} it is
not difficult to obtain equation for $n_{tr}(t)$:
\begin{equation}
  \label{eq:n-tr}
  \pdr{n_{tr}}=
(\gamma_c+q_c I)\,n_{cis}-\langle P_{tr}\rangle_{tr} n_{tr}\, ,
\end{equation}
where the angular brackets $\langle\ldots\rangle_\alpha$ 
stand for averaging over angles with the distribution function
$f_{\alpha}$ . Owing to the condition~\eqref{eq:norm}, this
equation does not depend on the form of the angular redistribution
probabilities.

The last square bracketed term on the right hand side of
Eq.~\eqref{eq:master} describes 
the process that equilibrates 
the side chain absorbing molecules and the polymer system
in the absence of irradiation.
The angular redistribution probabilities
$\Gamma_{\alpha-p}(\uvc{n},\uvc{n}')$ meet the normalization
condition, so that  thermal relaxation does not
change the total fractions $N_{tr}$ and $N_{cis}$. 
If there is no angular redistribution, 
then $\Gamma_{\alpha-p}(\uvc{n},\uvc{n}')=\delta(\uvc{n}-\uvc{n}')$
and both equilibrium angular distributions 
$f_{tr}^{(eq)}$ and $f_{cis}^{(eq)}$ are equal to $f_p$.

We can now derive the mean field models considered in
Refs.~\cite{Ped:1997,Ped:1998,Hvil:2001} by putting $\gamma_c=0$
(long-living \cis fragments), $\gamma_{cis}=\gamma_{tr}$ and assuming
that the angular redistribution probabilities
$\Gamma_{t-c}(\uvc{n},\uvc{n}')$ and $\Gamma_{c-t}(\uvc{n},\uvc{n}')$
are equal to the equilibrium distribution, $f_p=p(\uvc{n})$,
determined by the mean-field potential $V(\uvc{n})$:
$p(\uvc{n})\propto \exp(-V/k_B T)$.  In other words, this procedure
implies that the angular redistribution operators act as projectors
onto the angular distribution of polymer system.  This is the
order parameter dependent distribution that characterizes the
orientation of azo dye molecules after isomerization.

An alternative and a more general approach is to
determine the distribution function $f_p(\uvc{n},t)$
from the kinetic equation that can be written in the following form: 
\begin{align}
&
\pdr{f_{p}(\uvc{n},t)}= -\sum_{\alpha=\{tr,cis\}}
\gamma_p^{(\alpha)}\,n_{\alpha}(t)\,
\Bigl[\,f_p(\uvc{n},t) \notag\\
&
-\int\Gamma_{p-\alpha}(\uvc{n},\uvc{n}')
f_{\alpha}(\uvc{n}',t)\,\dd\uvc{n}'\,\Bigr]\,.
\label{eq:gen-p}
\end{align}
Equations for the angular distribution functions 
$f_{tr}(\uvc{n},t)$ and $f_{cis}(\uvc{n},t)$ can be derived
from Eq.~\eqref{eq:master} by using 
the relations~\eqref{eq:wc-t}--\eqref{eq:n-tr}.
The result is as follows
\begin{align}
& 
n_{cis}\,\pdr{f_{cis}}= 
-n_{tr}\,\Bigl[\,\langle P_{tr}\rangle_{tr} f_{cis} 
\notag\\
&
-
\int\Gamma_{t-c}(\uvc{n},\uvc{n}')P_{tr}(\uvc{n}')
f_{tr}(\uvc{n}',t)\,\dd\uvc{n}'\,\Bigr]\notag\\
&
-\gamma_{cis}\,n_{cis}\Bigl[\,f_{cis}
\notag\\
&
-\int\Gamma_{cis-p}(\uvc{n},\uvc{n}')
f_{p}(\uvc{n}',t)\,\dd\uvc{n}'\,\Bigr]
\label{eq:gen-cis}
\end{align}
\begin{align}
& 
n_{tr}\,\pdr{f_{tr}}= -n_{tr}
\left[ P_{tr}(\uvc{n})-\langle P_{tr}\rangle_{tr}
\right] f_{tr}
\notag\\
&
+\gamma_c\,n_{cis}
\int\Gamma_{c-t}^{(sp)}(\uvc{n},\uvc{n}')
f_{cis}(\uvc{n}',t)\,\dd\uvc{n}'\notag\\
&
+q_c I n_{cis}
\int\Gamma_{c-t}^{(ind)}(\uvc{n},\uvc{n}')
f_{cis}(\uvc{n}',t)\,\dd\uvc{n}'
\notag\\
&
-(\gamma_c+q_c I)\,n_{cis}\,f_{tr}
-\gamma_{tr}\,n_{tr}\Bigl[\, f_{tr}\notag\\
&
-\int\Gamma_{tr-p}(\uvc{n},\uvc{n}')
f_{p}(\uvc{n}',t)\,\dd\uvc{n}'\,\Bigr]\, .
\label{eq:gen-tr}
\end{align}
 
Eq.~\eqref{eq:n-tr} and
Eqs.~\eqref{eq:gen-p}--\eqref{eq:gen-tr}  
can be regarded as a starting point for the formulation
of a number of phenomenological models of POA.
We have already shown that these models include the mean field
theories considered in Refs.~\cite{Ped:1997,Ped:1998,Hvil:2001}.

An alternative model was suggested in
Refs.~\cite{Puch:1998,Puch:1999}, where angular distribution of \cis
isomers is assumed to be stationary and isotropic,
$f_{cis}=(4\pi)^{-1}$. This assumption implies that
\begin{equation}
  \label{eq:as-gamma}
\gamma_{cis}=\gamma_p^{(cis)}=0  
\end{equation}
and 
$\Gamma_{t-c}(\uvc{n},\uvc{n}')=f_{cis}$.
In order to ensure that $f_p=f_{tr}$ at the equilibrium state,
this model uses the relation
\begin{equation}
  \label{eq:as-p}
  \Gamma_{\alpha-p}(\uvc{n},\uvc{n}')
=\Gamma_{p-\alpha}(\uvc{n},\uvc{n}')=\delta(\uvc{n}-\uvc{n}')\,.
\end{equation}
So, after neglecting biaxiality we can 
obtain the results of Refs.~\cite{Puch:1998,Puch:1999} 
by additionally putting
$\Gamma_{c-t}^{(sp)}(\uvc{n},\uvc{n}')=f_{tr}(\uvc{n},t)$
and $\Gamma_{c-t}^{(ind)}(\uvc{n},\uvc{n}')=(4\pi)^{-1}$.

In this paper we consider another simple model by assuming that all the
angular redistribution operators $\Gamma_{t-c}$
and $\Gamma_{c-t}$ take the following isotropic form:
\begin{align}
&
 \Gamma_{c-t}^{(sp)}(\uvc{n},\uvc{n}')=
\Gamma_{c-t}^{(ind)}(\uvc{n},\uvc{n}')\notag\\
&
=\Gamma_{t-c}(\uvc{n},\uvc{n}')=\frac{1}{4\pi}\equiv f_{iso}\,.
\label{eq:as-iso}
\end{align}
Since we have neglected anisotropy of \cis fragments 
in Eq.~\eqref{eq:wt-c},
it is reasonable to suppose that the equilibrium
distribution of \cis molecules is also isotropic,
$f_{cis}^{(eq)}=f_{iso}$. From the other hand, 
the equilibrium angular distribution of \trans fragments is determined
by the polymer system: $f_{tr}^{(eq)}=f_{p}$.
So, we can use Eqs.~\eqref{eq:as-gamma}--\eqref{eq:as-iso}
to derive the system of kinetic equations from 
Eqs.~\eqref{eq:n-tr}--\eqref{eq:gen-tr}. The result is given by
\begin{subequations}
\label{eq:mod}
\begin{align}
& n_{cis}\pdr{f_{cis}}= 
n_{tr}\langle P_{tr}\rangle_{tr}\,(\,f_{iso} - f_{cis}\,)\,, 
\label{eq:mod-cis}\\
& 
n_{tr}\pdr{f_{tr}}=\left(
\langle P_{tr}\rangle_{tr}-P_{tr}
\right)n_{tr} f_{tr}+
(\gamma_c + q_c I)\notag\\ 
&
\times n_{cis} (\,f_{iso}-f_{tr}\,)+
\gamma_{tr} n_{tr} (\,f_p-f_{tr}\,) ,
\label{eq:mod-tr}\\
&\pdr{f_{p}}=\gamma_p\, n_{tr}(f_{tr}-f_p)\, , 
\label{eq:mod-p}
\end{align}
\end{subequations}
where $\gamma_p\equiv\gamma_p^{(tr)}$.

At this stage we have the equation for the fractions~\eqref{eq:n-tr}
and Eqs.~\eqref{eq:mod} for the angular distribution functions. 
It is now our task to describe the temporal evolution of photoinduced
anisotropy in terms of 
the averaged components of the order parameter tensor~\cite{Gennes:bk:1993}
\begin{equation}
  \label{eq:tens}
  S_{ij}(\uvc{n})=2^{-1}\,(3 n_i n_j -\delta_{ij})\, .
\end{equation}

The simplest case occurs for the order parameters of \cis molecules.
Integrating Eq.~\eqref{eq:mod-cis} multiplied by $S_{ij}(\uvc{n})$
over angles gives the following result:
\begin{equation}
  \label{eq:ord-cis}
  \pdr{S_{ij}^{(cis)}}=-n_{tr}\langle P_{tr}\rangle_{tr}\ S_{ij}^{(cis)}\, ,
\end{equation}
where $ S_{ij}^{(\alpha)}\equiv\langle S_{ij}(\uvc{n})\rangle_\alpha$. 
Rotational diffusion of \cis molecules can be taken into account by
replacing the coefficient on the right
hand side of Eq.~\eqref{eq:ord-cis} with
$n_{tr}\langle P_{tr}\rangle_{tr}+ 6 D_r$, where
$D_r$ is the rotational diffusion constant. In any case, 
ordering kinetics of \cis molecules is irrelevant provided
the subsystem of \cis groups is initially isotropic.
The reason is that 
photo-isomerization does not make angular distribution of \cis
fragments anisotropic in our model. 

We can now multiply Eqs.~\eqref{eq:mod-tr} and~\eqref{eq:mod-p}
by  $S_{ij}(\uvc{n})$ and integrate the resulting equations over
the angles to obtain the following system: 
\begin{subequations}
\label{eq:gen-ord}
\begin{align}
&
n_{tr}\pdr{S_{ij}^{(tr)}}= -2/3\, q_t I u\,
n_{tr}\,G_{ij;\,xx}^{(tr)}-n_{cis}\notag\\ 
&
\times (\gamma_c+q_c I)\,S_{ij}^{(tr)} +
\gamma_{tr} n_{tr} (S_{ij}^{(p)}-S_{ij}^{(tr)})\, ,
\label{eq:gen-ord-tr}\\
&\pdr{S_{ij}^{(p)}}=-\gamma_{p} n_{tr}(S_{ij}^{(p)}-S_{ij}^{(tr)})\, ,
\label{eq:gen-ord-p}
\end{align}
\end{subequations}
where $G_{ij;\,mn}^{(\alpha)}$ is the order parameter
correlation function (correlator) defined as follows:
\begin{equation}
  \label{eq:cor-fun}
  G_{ij;\,mn}^{(\alpha)}=
\langle S_{ij}(\uvc{n})S_{mn}(\uvc{n}) \rangle_\alpha-
S_{ij}^{(\alpha)}\,S_{mn}^{(\alpha)}\, .
\end{equation}
These functions characterize response of the side groups to the
pumping light.

Eqs.~\eqref{eq:gen-ord} will give the system for the components of the
order parameter tensor, if a closure can be found linking the
correlation functions and $S_{ij}^{(\alpha)}$.  The simplest closure
can be obtained by writing the products of the order parameter
components as a sum of spherical harmonics and neglecting the high
order harmonics.  In particular, applying this procedure to the
diagonal order parameter components gives the following parabolic
approximation for the order parameter
autocorrelators~\cite{Kis:epj:2001}
\begin{equation}
  \label{eq:decoup}
  G_{ii;\,ii}^{(tr)}\approx
1/5+2/7\,\langle S_{ii}\rangle_{tr}-\langle S_{ii}\rangle_{tr}^2\, .
\end{equation}

According to the definition~\eqref{eq:cor-fun}, the autocorrelators
$G_{ii;\,ii}^{(tr)}$ must be non-negative. 
From the other hand,
the values of $\langle S_{ii}\rangle_{tr}$
are ranged from $-0.5$ to $1$~\cite{Gennes:bk:1993} and
the expression on the right hand side of Eq.~\eqref{eq:decoup}
takes negative values on the specified interval. In order to 
restore the correct behavior, 
we need to modify the approximation~\eqref{eq:decoup}.
Our assumption is that it can be done by rescaling the
order parameter components: $\langle S_{ii}\rangle_{tr}\to
\lambda\langle S_{ii}\rangle_{tr}$, where $\lambda$ is the coefficient  
that takes into
account contributions coming from the high order harmonics.
In our calculations we used the largest value of $\lambda$,
$\lambda_{max}=(1+0.6\sqrt{30})/7$.
This implies that there are
no fluctuations provided the molecules are perfectly aligned 
along the coordinate unit vector $\uvc{e}_i$ and
$G_{ii;\,ii}^{(tr)}=0$ at
$\langle S_{ii}\rangle_{tr}=1$.

By using the modified parabolic approximation for the correlators,
we can now write down  
the resulting system for the diagonal components of
the order parameter tensor in the final form:
\begin{align}
&n_{tr}\pdr{S}= -2u/3\, q_t I (5/7+2\lambda/7\,S-\lambda^2 S^2) n_{tr}\notag\\
&-(\gamma_c+q_c I)\, n_{cis} S +\gamma_{tr} n_{tr} (S_p-S),
\label{eq:ord-tr1}
\end{align}
\begin{align}
&n_{tr}\pdr{\Delta S}= 2u/3\, q_t I\lambda (2/7+\lambda S) n_{tr}\Delta S 
-n_{cis}\notag\\
&\times (\gamma_c+q_c I)\Delta S +
\gamma_{tr} n_{tr} (\Delta S_p-\Delta S),
\label{eq:ord-tr2}
\end{align}
\begin{align}
&\pdr{S_p}=-\gamma_{p} n_{tr} (S_p-S)\, ,
\label{eq:ord-p1}
\end{align}
\begin{align}
&\pdr{\Delta S_p}=-\gamma_{p} n_{tr} (\Delta S_p-\Delta S)\, ,
\label{eq:ord-p2}
\end{align}
where $S\equiv \langle S_{xx}\rangle_{tr} $,
$\Delta S\equiv \langle S_{yy}-S_{zz}\rangle_{tr} $,
$S_p\equiv \langle S_{xx}\rangle_{p} $ and
$\Delta S_p\equiv \langle S_{yy}-S_{zz}\rangle_{p} $.

\begin{figure*}[!tbh]
%\vskip5mm
\centering
\resizebox{150mm}{!}{\includegraphics*{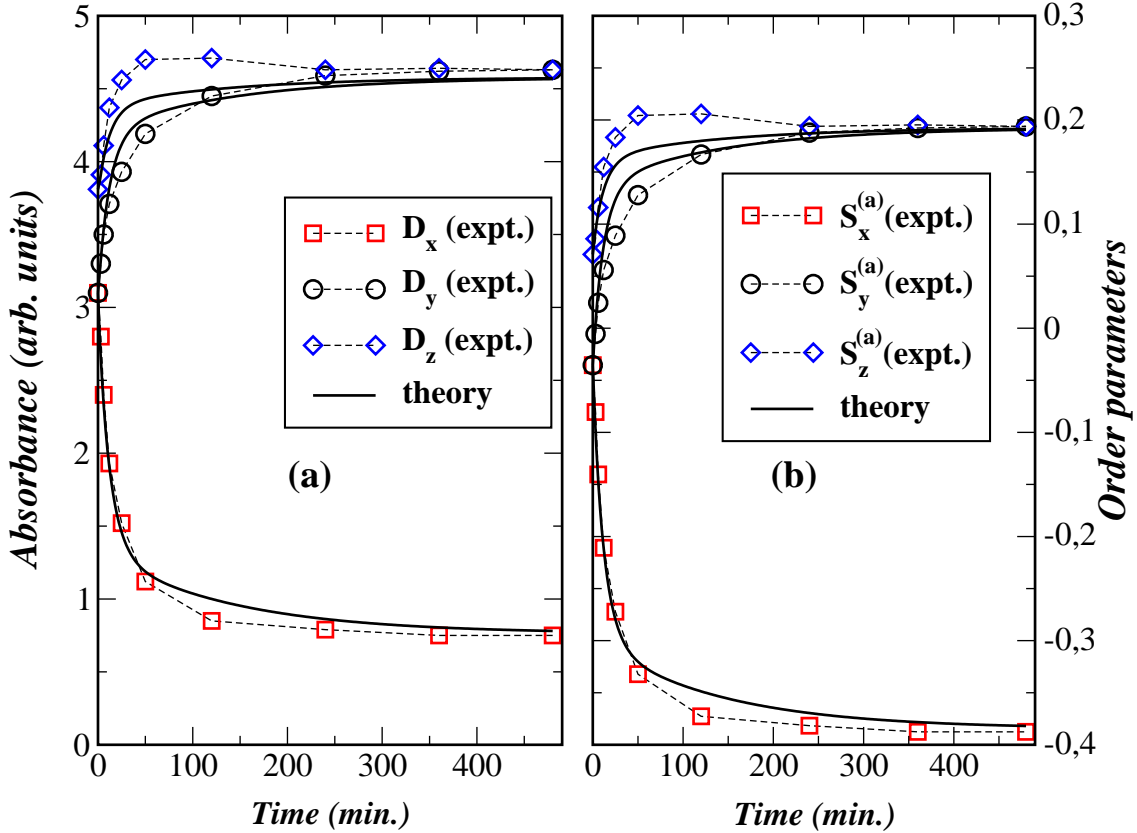}}
\caption{%
Dependencies of (a) the principal absorption coefficients and 
(b) the absorption order parameters on the irradiation time
at $\lambda_{ex}=488$~nm and $I=2$~W/cm$^2$.
Theoretical curves for the order parameters 
and the absorption coefficients
are calculated at $\gamma_c=0.0$,
$r\equiv q_c/q_t=60.0$, $q_t I=0.01$ and $u=38.6$. 
}
\label{fig:P2-488}
\end{figure*}

\begin{figure*}[!tbh]
%\vskip5mm
\centering
\resizebox{150mm}{!}{\includegraphics*{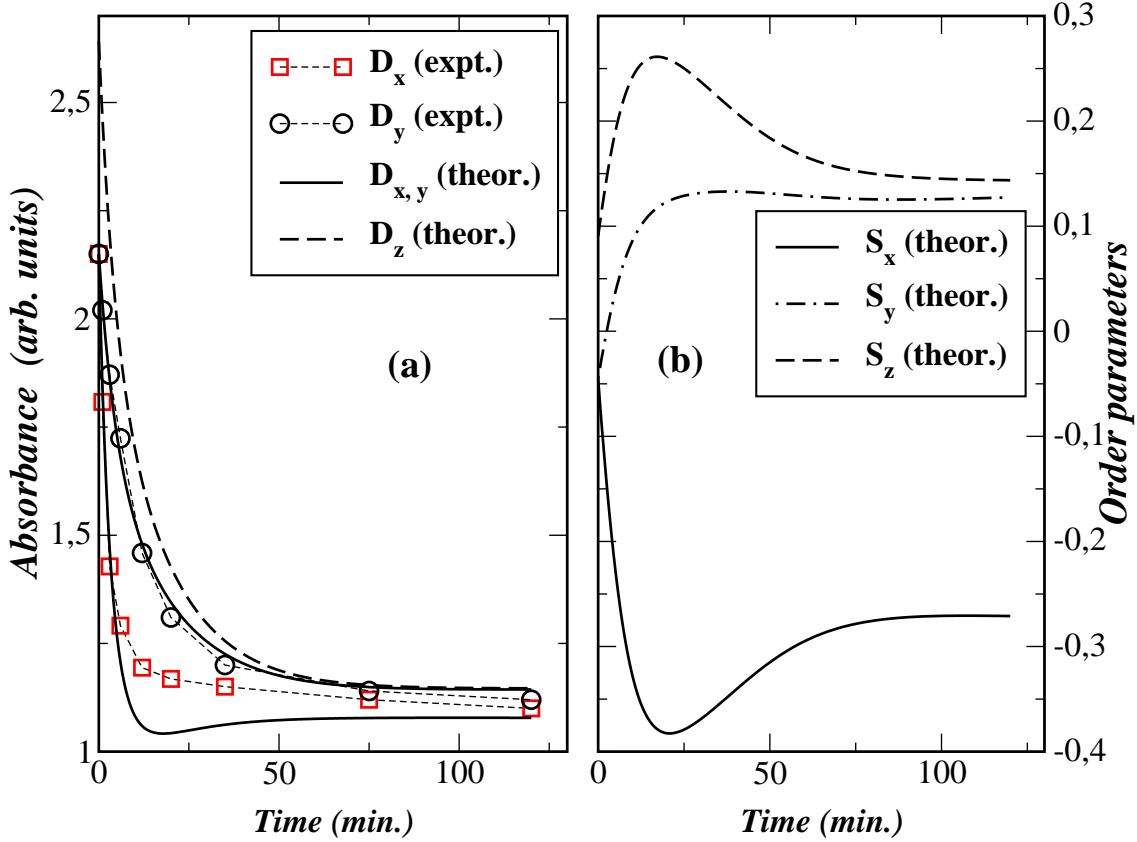}}
\caption{%
Dependencies of (a) the principal absorption coefficients and 
(b) the order parameter components on the irradiation time
at $\lambda_{ex}=365$~nm and $I=3$~mW/cm$^2$.
Theoretical curves for the diagonal components
of the order parameter tensor 
and the absorption coefficients
are calculated at $\gamma_c=0.0$,
$r\equiv q_c/q_t=0.1$, $q_t I =0.06$ and $u=8.4$. 
}
\label{fig:P2-365}
\end{figure*}

\section{Numerical results}
\label{sec:num-res}

In this section we demonstrate how our model can be employed to
interpret the experimental data of the UV absorption measurements for
different irradiation doses.  For this purpose we will use the data
obtained by Dr. O.~Yaroshchuk and co-workers (Institute of Physics of
NASU, Ukraine).  Since the experimental procedure has been described
in Ref.~\cite{Kis:epj:2001} and a more comprehensive study is the
subject of joint publication~\cite{Kis:pre:unpb}, only a brief summary
will be given below.

According to Ref.~\cite{Kis:epj:2001}, the normally incident pumping
UV light used in the experiments is propagating along the $z$-axis and
is linearly polarized with the polarization vector $\vc{E}$ parallel
to the $x$-axis. The irradiation was provided in several steps
followed by absorption measurements after the waiting time period
taken to be longer than 15~min. These measurements were carried out to
yield the optical density components $D_x$ and $D_y$ for the testing
beam which is linearly polarized along the $x$ and $y$ axes,
respectively. The wavelength of the testing light was tuned to the
absorption maximum of azobenzene chromophores at $\lambda_t=343$~nm.
The principal absorption coefficients $D_i$ can be related to the
concentrations and the order parameters as follows
\begin{align}
  &
  D_i\propto \langle\sigma_{ii}^{(tr)}\rangle_{tr}\,n_{tr}
+\sigma^{(cis)}\, n_{cis}\notag\\
&
\propto (1+u^{(a)}(2\,S_i+1)/3) n_{tr}+q_{ct} n_{cis}\,,
\label{eq:abs_coef}
\end{align}
where $S_i\equiv \langle S_{ii}\rangle_{tr}$;
$u^{(a)}$ is the absorption anisotropy parameter and 
$q_{ct}$ is the ratio of $\sigma^{(cis)}$ and
$\sigma_{\perp}^{(tr)}$ at the wavelength of probing light.

The lifetime of the \cis fragments was found to be much longer than
the periods examined. So, we can safely
take the limit of long-living \cis molecules and 
neglect $\gamma_c$ in our calculations. 
In this case the photoselection mechanism discussed in
Sec.~\ref{sec:intro} can be thought to 
dominate the process of photo-reorientation.
There are, however, two sets of experimental data  
measured at two different wavelengths of the pumping UV light: 
$\lambda_{ex}=488$~nm and $\lambda_{ex}=365$~nm.
These data are shown in Fig.~\ref{fig:P2-488}a and 
Fig.~\ref{fig:P2-365}a, respectively. 

Fig.~\ref{fig:P2-488}a presents the case in which the wavelength of
light is far from the absorption maximum and dependencies of the
absorption coefficients $D_x$ and $D_y$ on the irradiation time are
typical of photoorientation mechanism. In this case the fraction of
\cis molecules is negligible and a sum of all the principal absorption
coefficients, $D_{tot}=D_x+D_y+D_z$, does not depend on irradiation
doses. For the photosteady state, where $D_y^{(st)}=D_z^{(st)}$, this
will yield the relation
\begin{equation}
  \label{eq:D_tot}
  D_{tot}=D_x^{(st)}+2 D_y^{(st)}=
D_x+D_y+D_z\,.
\end{equation}
The absorption component $D_z$ depicted in Fig.~\ref{fig:P2-488}a
was estimated by using Eq.~\eqref{eq:D_tot}.
In addition, Fig.~\ref{fig:P2-488}b shows the absorption
order parameters $S_i^{(a)}$ computed from the expression
\begin{equation}
  \label{eq:S_abs}
  S_i^{(a)}=\frac{2 D_i - D_j - D_k}{2(D_x+D_y+D_z)},\quad
i\ne j\ne k\,.
\end{equation}
From Eq.~\eqref{eq:abs_coef} with $n_{cis}\approx 0$ these
order parameters are proportional to $S_i$:
$S_i^{(a)}=u^{(a)}/(3+u^{(a)}) S_i$.

By contrast, referring to Fig.~\ref{fig:P2-365}a, it is seen that
both experimental dependencies $D_x$ and $D_y$
are decreasing functions of the irradiation time
under the wavelength is near the maximum of absorption band
with $\lambda_{ex}=365$~nm. It indicates 
that in this case 
POA is governed by the mechanism of photoselection.
So, we have the process of photo-reorientation 
characterized by two different mechanisms depending on the wavelength
of pumping light.

In order to characterize the regime of POA,
we can use the fraction of \cis fragments in the
photo-stationary state.
From Eq.~\eqref{eq:n-tr} this fraction is given by
\begin{equation}
n_{cis}^{(st)}=\frac{3+u(1+2 S_{st})}{3(r+1)+u(1+2 S_{st})}\,,
  \label{eq:nc_st}
\end{equation}
where $r\equiv (\gamma_c+q_c I)/(q_t I)$,
$S_{st}\equiv S_x^{(st)}$ and
the corresponding value of the order parameter is a solution
of the following equation  
\begin{align}
&
2 u\,(1/5\,+2\lambda/7\, S_{st}-\lambda^2 S_{st}^{\,2})\notag\\
&
=-S_{st}(3+u(1+2 S_{st}))\,,
  \label{eq:S_st}
\end{align}
deduced by using Eqs.~\eqref{eq:ord-tr1}
and~\eqref{eq:ord-p1}. 

When $\gamma_c=0$, the parameter $r$ in Eq.~\eqref{eq:nc_st} is the
ratio of $q_c$ and $q_t$.  At small values of $r$ it will yield the
fraction $n_{cis}^{(st)}$ that is close to the unity and we have the
kinetics of POA in the regime of photoselection.  In the opposite case
of sufficiently large values of $r$ the photosteady fraction of \cis
molecules will be very small that is typical of the photoorientation
mechanism.  This effect is demonstrated in Figs.~\ref{fig:P2-488}
and~\ref{fig:P2-365}. The figures show theoretical curves computed at
different values of $r$: $60.0$ and $0.1$, respectively.  From
Eqs.~\eqref{eq:nc_st} and~\eqref{eq:S_st} the corresponding fractions
of \cis fragments in the photostationary state can be estimated at
about $0.02$ and $0.96$.

The theoretical curves are computed by solving the rate equations
deduced in the previous section.  Initial values of the order
parameters $S(0)$ and $\Delta S(0)$ are taken from the experimental
data measured at $\lambda_{ex}=488$~nm.  Since the system is initially
at the equilibrium state, the remaining part of the initial conditions
is: $S_p(0)=S(0)$, $\Delta S_p(0)=\Delta S(0)$, $n_{tr}(0)=1$ and
$n_{cis}(0)=0$.

Numerical calculations in the presence of irradiation were followed by
computing the stationary values of $S$ and $\Delta S$ to which the
order parameters decay after switching off the irradiation at time
$t$.  The experimental estimate for the relaxation time characterizing
decay of $D_{i}(t)$ to its stationary value after switching off the
irradiation is about 1~h.  The theoretical value of this relaxation
time, deduced from solution of the kinetic equations in the absence of
irradiation, is $1/(\gamma_p + \gamma_{tr})$.  So, in the simplest
case, we can assume both relaxation times, $\tau_p$
($\gamma_p=1/\tau_p$) and $\tau_{tr}$ ($\gamma_{tr}=1/\tau_{tr}$), to
be equal 120~min.

\begin{table}[htbp]
\begin{ruledtabular}
  \begin{tabular}{cD{.}{.}{-1}D{.}{.}{-1}}
$\lambda_{ex}$ (nm) &365 & 488 \\
\colrule
$I$ (W/cm$^2$) & 0.003 & 2\\
$\sigma_{cis}$ $\times 10^{-18}$(cm$^2$) & 15 & 3.14\\
$\sigma^{(tr)}/\sigma_{cis} $ & 57.2 & 0.5\\
$\sigma_{||}^{(tr)}/\sigma_{\perp}^{(tr)}$ & 9.4 & 39.6\\
$\Phi_{cis\to trans}$ (\%) & 15 & 10\\
$\Phi_{trans\to cis}$ (\%) & 10 & 5\\ 
  \end{tabular}
\end{ruledtabular}
  \caption{Photochemical parameters.}
  \label{tab:param}
\end{table}

There are a number of additional photochemical parameters that enter
the model and are listed in Table~\ref{tab:param}.
The table shows the estimates for
absorption cross section of \cis molecules 
$\sigma^{(cis)}$ and 
average absorption cross section of \trans fragments,
$\sigma^{(tr)}=(\sigma_{||}^{(tr)}+2\sigma_{\perp}^{(tr)})/3$,
obtained from the UV spectra of the polymer solved in toluene.
We shall omit details on
the method of evaluation which
is briefly described in Ref.~\cite{Kis:epj:2001}.

For this polymer the absorption anisotropy parameters and the quantum
efficiencies are unknown and need to be fitted.  We used the value of
$S_{st}$ as an adjustable parameter, so that the anisotropy parameters
$u$ and $u^{(a)}$ can be derived from Eq.~\eqref{eq:S_st} and from the
experimental value of the absorption order parameter $S_{st}^{(a)}$
measured at $\lambda_{ex}=488$~nm in the photosteady state.  The
numerical results presented in Figs.~\ref{fig:P2-488}
and~\ref{fig:P2-365} are computed at $u^{(a)}=11.0$ and $q_{ct}=2.15$.
Note that the quantum efficiencies are of the same order of magnitude
as the experimental values for other azobenzene
compounds~\cite{Mita:1989}.

\section{Discussion}
\label{sec:concl}

Despite our theoretical considerations are rather phenomenological
they emphasize the key points that should be addressed by such kind of
theories. These are the angular redistribution probabilities and the
order parameter correlation functions.  The redistribution operators,
in particular, define how the system relaxes after switching off the
irradiation and can serve to introduce self-consistent fields.  The
correlators describe response of \trans molecules to the exciting
light and determine the properties of the photosaturated state.

Our simple model relies on the assumption that the \cis fragments are
isotropic and do not affect the ordering kinetics directly.  
By contrast to the mean field models of
Refs.~\cite{Ped:1997,Ped:1998,Hvil:2001},
it implies that the presence of long-living
angular correlations is irrelevant for \cis molecules. Certainly, this
is the simplest case to start from before studying more complicated
models. The model depends on a few parameters that enter the equations
and that can be estimated from the experimental data.  Only the
absorption anisotropy parameters and the quantum yields need to be
adjusted.

After making comparison between the experimental data and the
theoretical results we can conclude that the theory correctly captures
the basic features of POA in azopolymers.  It predicts that in the
limit of long-living \cis fragments the regime of POA will be governed
by the parameter $r$ which is the ratio of $\Phi_{cis\to
  trans}\sigma^{(cis)}$ and $\Phi_{trans\to
  cis}\sigma_{\perp}^{(tr)}$.

For large values of $r$ the fraction of \cis molecules in the
photosaturated state is negligible  and dependencies
of the principal absorption coefficients on illumination doses
are typical for the kinetics in the regime of photoorientation.
It means that the $\cis\to\trans$ 
transitions stimulated by the exciting light
will efficiently deplete the \cis state and the absorption
coefficients are controlled by the terms
proportional to the order parameter of \trans molecules
(see Eq.~\eqref{eq:abs_coef}).

By contrast, if the value of $r$ is sufficiently small,
there is nothing to prevent the \cis state from being populated
under the action of UV light and the fraction of \cis molecules
approaches the unity upon illumination. In this case 
the contribution of \trans isomers to the absorbance becomes
negligible as the illumination dose increases and we have the
kinetics of POA dominated by the photoselection mechanism.

We have thus demonstrated that the predictions of the theory
are in accord with the experimental data, where
the difference in the photochemical parameters at $\lambda_{ex}=365$~nm and 
$\lambda_{ex}=488$~nm can be attributed to the interplay
between $\pi\pi^{*}$ and n$\pi^{*}$ transitions of
the azobenzene moieties. Our findings confirm the
conclusion that the dependence of photoisomerization rate on molecular
axes orientation (see Eqs.~\eqref{eq:wc-t} and~\eqref{eq:ptr}) plays a
leading part in the process of photo-reorientation.

Our concluding remark concerns an overall strategy used throughout
this paper. We have analyzed the general structure of phenomenological
models to find out how the models incorporate different physical
assumptions. Then we studied the simple model and compared the results
of calculations with the experimental data.  Recently we have applied
this strategy to describe the photoorientation regime of POA in the
polymer, where the out-of-plane reorientation of the dye molecules is
suppressed~\cite{Kis:epj:2001}.  Since the method is found to be
applicable to the different regimes of POA, we have a useful tool for
studying photoinduced ordering processes in azopolymers.  But, not
speaking of the parabolic approximation for the correlators, even the
very possibility to describe the kinetics in terms of one-particle
distribution functions needs to be justified by a more detailed
theory. The structure of models is likely to be recovered in the mean
field approximation~\cite{Luben:bk:1995} and we hope that this work
will stimulate further progress in this direction.

\begin{acknowledgments}
We are grateful to Dr. O.~Yaroshchuk 
(Institute of Physics of NASU, Ukraine)
for stimulating discussions
and his kind permission to use the experimental data. 
\end{acknowledgments}

\bibliographystyle{apsrev}
\bibliography{polymer,scatter,lc,quant}

\end{document}